\documentclass[10pt,twocolumn,letterpaper]{article}
\usepackage{iccv}

\usepackage[accsupp]{axessibility}

\definecolor{iccvblue}{rgb}{0.21,0.49,0.74}

\usepackage{times}
\usepackage{epsfig}
\usepackage{graphicx}
\usepackage{amsmath}
\usepackage{amssymb}
\usepackage{float}
\usepackage{booktabs}
\usepackage{multirow}
\usepackage{pifont}
\usepackage{tikz}
\usepackage{url}
\usepackage{enumitem}

\usepackage{colortbl}

\usepackage[pagebackref,breaklinks,colorlinks,allcolors=iccvblue]{hyperref}

\def\paperID{15719}
\def\confName{ICCV}
\def\confYear{2025}

\title{Align Your Rhythm: Generating Highly Aligned Dance Poses with Gating-Enhanced Rhythm-Aware Feature Representation}

\author{\textbf{Congyi Fan$^1$},
\textbf{Jian Guan$^{1}$\thanks{Corresponding author: j.guan@hrbeu.edu.cn}}, 
\textbf{Xuanjia Zhao$^1$},
\textbf{Dongli Xu$^2$}, \\
\textbf{Youtian Lin$^3$},
\textbf{Tong Ye$^1$},
\textbf{Pengming Feng$^4$},
\textbf{Haiwei Pan$^1$}
\\ \\
$^1$College of Computer Science and Technology, Harbin Engineering University \\
$^2$Shanghai Academy of Artificial Intelligence for Science \\
$^3$School of Intelligence Science and Technology, Nanjing University \\
$^4$State Key Laboratory of Space-Ground Integrated Information Technology \\
}

\begin{document}

\maketitle

\begin{abstract}

Automatically generating natural, diverse, and rhythmic human dance movements driven by music is vital for virtual reality and film industries. However, generating dance that naturally follows music remains a challenge, as existing methods lack proper beat alignment and exhibit unnatural motion dynamics. In this paper, we propose Danceba, a novel framework that leverages gating mechanism to enhance rhythm-aware feature representation for music-driven dance generation, which achieves highly aligned dance poses with enhanced rhythmic sensitivity. Specifically, we introduce Phase-Based Rhythm Extraction (PRE) to precisely extract rhythmic information from musical phase data, capitalizing on the intrinsic periodicity and temporal structures of music. Additionally, we propose Temporal-Gated Causal Attention (TGCA) to focus on global rhythmic features, ensuring that dance movements closely follow the musical rhythm. We also introduce Parallel Mamba Motion Modeling (PMMM) architecture to separately model upper and lower body motions along with musical features, thereby improving the naturalness and diversity of generated dance movements. Extensive experiments confirm that Danceba outperforms state-of-the-art methods, achieving improved rhythmic alignment and motion diversity. Project page: \href{https://danceba.github.io/}{https://danceba.github.io/}. 

\end{abstract}

\section{Introduction}
\label{sec:intro}

Music-driven dance generation aims to automatically synthesize realistic and expressive 3D dance movements synchronized precisely with a given musical sequence, enabling natural animation of 3D character models. Due to its broad applications in gaming, virtual reality, film production, and interactive entertainment~\cite{zhu2023human, 10.1145/2735951, 10.1145/2702613.2732759, tsuchida2024dance}, this task has attracted considerable attention. However, generating dances that accurately reflect musical rhythm while preserving natural and diverse movement dynamics remains a significant challenge, as it requires effectively capturing and utilizing complex temporal relationships between music and motion.

Recently, generative models have demonstrated remarkable success across various tasks, such as text and image generation~\cite{radford2019language,ho2020denoising}. Similarly, these generative approaches have also been applied to music-driven dance generation tasks~\cite{siyao2022bailando, zhuang2023gtn, siyao2023bailando++, huang2024enhancing, wang2024explore, tseng2023edge, 10.1145/3581783.3612307, li2023finedance, li2024lodge, zheng2024beat}. 
Among these, Bailando~\cite{siyao2022bailando} introduces a cross-conditional causal attention mechanism to align musical cues with corresponding dance poses, significantly improving the quality of generated dance movements. However, its attention mechanism primarily captures local correspondence between music and poses, failing to sufficiently utilize  global rhythmic characteristics within music, thus limiting the expressiveness and rhythmic coherence of the generated dances.

Follow-up studies~\cite{zhuang2023gtn, siyao2023bailando++, huang2024enhancing, wang2024explore} have extended Bailando’s design to further improve dance generation performance. For instance, Enhancing-Bailando~\cite{huang2024enhancing} employs a pretrained audio encoder (MERT)~\cite{li2023mert} to obtain richer musical feature representations. However, this encoder primarily captures semantic information from general audio data rather than explicitly modeling rhythmic characteristics, limiting its effectiveness in precise beat-dance synchronization.  Moreover, Bailando and related methods typically concatenate three distinct input modalities, i.e., upper-body pose, lower-body pose, and musical features, and rely heavily on cross-conditional causal attention to align these inputs. Nevertheless, this attention mechanism primarily aligns pose and music at a local feature level, but struggles to capture continuous motion dynamics, limiting the naturalness and diversity of generated dance movements.

To further tackle the rhythm alignment issue, Beat-It~\cite{zheng2024beat} explicitly introduces beat synchronization via a beat distance predictor. However, such an explicit synchronization approach imposes rigid control constraints and strong regularization during generation, significantly limiting the diversity and expressiveness of generated dances~\cite{10.1145/3581783.3612307}. Thus, a more balanced approach 
capable of achieving precise rhythmic alignment without sacrificing motion naturalness and diversity is highly desirable.

To this end, we propose a novel framework equipped with a gating-enhanced rhythm-aware feature representation for music-driven dance generation, i.e., Danceba, which achieves highly aligned dance poses with enhanced rhythmic sensitivity. Specifically, Danceba introduces Phase-Based Rhythm Extraction (PRE), a rhythm-aware feature extraction module that explicitly captures rhythmic characteristics from musical phase information. Leveraging the temporal phase information intrinsic to rhythm perception, PRE effectively separates rhythmic from semantic music features, significantly improving the rhythmic alignment between generated dance sequences and musical inputs.

To further enhance beat-dance alignment\footnote{Note that, the terms ``beat'' and ``rhythm'' are used interchangeably in music processing following DiffDance~\cite{10.1145/3581783.3612307} and Bailando++~\cite{siyao2023bailando++}.} using the rhythm-aware features extracted by PRE, it is essential to establish a stronger contextual association between musical rhythm and dance movements. To achieve this, we present a Temporal-Gated Causal Attention (TGCA) mechanism, enhanced by a gating strategy to explicitly prioritize and leverage global rhythmic structures within cross-conditional causal attention. By integrating gating mechanism, TGCA effectively strengthens the contextual connection between music rhythm and generated movements, ensuring that generated dance sequences accurately reflect the underlying musical rhythm and exhibit greater coherence and expressive diversity.

Furthermore, to overcome the limitations of traditional attention-based approaches in motion modeling, we introduce Parallel Mamba Motion Modeling (PMMM), inspired by recent advances demonstrating that Mamba architectures achieve state-of-the-art performance in modeling complex human motion sequences~\cite{zhang2025motion, zeng2024light, FTMoMamba}. Our parallel Mamba architecture separately models upper and lower-body dance sequences, with input features enhanced by Temporal-Gated Causal Attention (TGCA), which incorporates rhythm-aware representations alongside distinct upper and lower-body pose features. This design effectively captures diverse and nuanced motion characteristics. By leveraging Mamba’s superior sequential modeling capability, our method significantly improves the expressiveness, diversity, and temporal coherence of generated dance movements. Finally, an additional TGCA module is applied after PMMM to further reinforce rhythmic alignment, ensuring that the generated dances precisely reflect the musical rhythm.

Extensive experiments show that our method improves $\text{FID}_k$ by 48.68\%, $\text{Div}_k$ by 7.0\%, $\text{Div}_g$ by 16.3\%, and BAS by 12.0\% over the second-best results, while maintaining a competitive $\text{FID}_g$ of 11.90. This significantly surpasses state-of-the-art methods, producing dance movements that are more natural, diverse, and better synchronized with musical rhythm.

\section{Related Work}
\subsection{Music-Driven Dance Generation}

Music-driven dance generation focuses on automatically synthesizing 3D dance movements from musical input, creating temporally coherent pose sequences for animating 3D character models. It has attracted increasing attention in both the computer vision and virtual avatar communities~\cite{kim2022brand, alexanderson2023listen} due to its applications in entertainment, gaming, and virtual reality. Early approaches, including~\cite{holden2016deep, lee2019dancing, sun2020deepdance, li2020learning, 10.1145/3485664, huang2021, li2021ai}, investigated diverse methodologies for this task. The study in \cite{li2020learning} utilized a standard Transformer~\cite{vaswani2017attention} with a late-fusion module to predict subsequent dance steps. 
The FACT framework~\cite{li2021ai} introduced the AIST++ dataset and employed a cross-modal Transformer with full attention mechanisms for music-to-dance generation. However, these initial methods neglect the rhythm features of music and struggle to effectively integrate musical and motion features, leading to suboptimal generation quality. 

Bailando~\cite{siyao2022bailando} reformulated dance generation as a next-step pose prediction problem using cross-conditional causal attention, achieving high-quality music-driven dance synthesis. Subsequent works~\cite{zhuang2023gtn, siyao2023bailando++, huang2024enhancing} extended this framework:   Bailando++~\cite{siyao2023bailando++} enhanced pattern recognition with musical context encoding, while Enhancing-Bailando~\cite{huang2024enhancing} incorporated a pre-trained MERT audio encoder~\cite{li2023mert} for improved feature extraction.  However, these methods often overlook rhythmic structures or rely on generic audio features. Beat-It~\cite{zheng2024beat} introduced explicit beat alignment via a conditional diffusion model~\cite{ho2020denoising, nichol2021glide}, ensuring precise synchronization. Yet, its reliance on rigid beat control signals and strong regularization may constrain pose diversity~\cite{10.1145/3581783.3612307}. 
In contrast, we emphasize rhythm-aware feature representation, achieving superior rhythmic alignment without sacrificing movement diversity.

\begin{figure*}[ht!]
    \centering
    \includegraphics[width=1\linewidth]{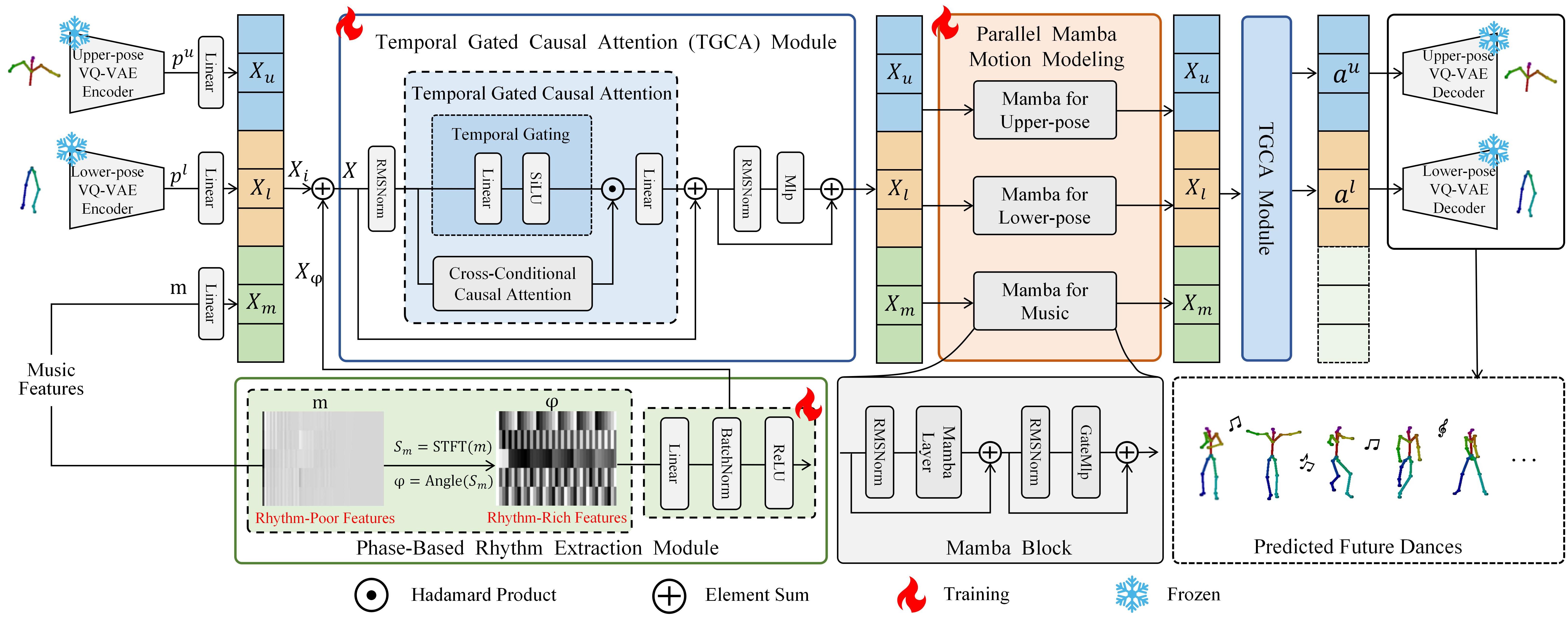}
    \caption{The overall framework of Danceba consists of three core modules: Phase-based Rhythm Extraction (PRE), Temporal-Gated Causal Attention (TGCA), and Parallel Mamba Motion Modeling (PMMM). PRE precisely extracts rhythm-aware features from musical phase information, providing accurate rhythmic signals. These rhythm-enhanced music features are fused with pose embeddings (upper-body and lower-body poses) and processed through the TGCA module, which utilizes gating mechanism to reinforce rhythmic sensitivity and align dance movements accurately with music beats. Parallel Mamba Motion Modeling separately models upper and lower body motion sequences, effectively capturing distinct dance dynamics aligned with rhythm-aware music features, thus significantly enhancing the naturalness, diversity, and temporal coherence of the generated dance movements. }
   \vspace{-14pt}
    \label{fig:pipeline}
\end{figure*}

\subsection{Mamba in Human Motion Generation}

The integration of Mamba~\cite{gu2023mamba} and Mamba2~\cite{dao2024transformers} into human motion generation represents a pivotal advancement, leveraging their efficient sequence modeling capabilities.  Within human motion generation, Mamba surpasses conventional Transformers in performance, as demonstrated by Motion Mamba~\cite{zhang2025motion}, which highlights its efficacy for extended motion sequence synthesis. Furthermore, InfiniMotion~\cite{InfiniMotion} utilizes Mamba's hierarchical and bidirectional temporal processing to generate continuous, smooth motion sequences of indefinite length. Recent approaches including SMCD~\cite{qian2024smcd}, Light-T2M~\cite{zeng2024light}, FTMoMamba~\cite{FTMoMamba}, and MMD~\cite{MMD} have further incorporated Mamba architectures, advancing the field through enhanced efficiency and high-fidelity motion modeling. 
Building upon recent advancements, Mamba has achieved state-of-the-art performance in human motion modeling~\cite{zhang2025motion, zeng2024light, FTMoMamba}, motivating our selection of Mamba over self-attention mechanisms for information exchange within dance sequences, which significantly enhances the diversity of generated dance motions.

\section{Proposed Method}

This section presents our proposed Danceba framework for music-driven 3D dance generation. By leveraging gating-enhanced rhythm-aware features, Danceba ensures precise alignment between dance poses and music. The overall framework is illustrated in Figure~\ref{fig:pipeline}, and its components are detailed as follows: Section~\ref{sec:Preliminary} outlines the preliminary framework and key notations. Specifically, we introduce phase-based rhythm feature extraction (Section~\ref{sec:Rhythm Feature Extraction}), global beat-dance alignment via temporal-gated causal attention (Section~\ref{sec:TGCA}), and Mamba-based parallel motion modeling (Section~\ref{sec:Mamba-Based Motion Modeling}).

\subsection{Preliminary}
\label{sec:Preliminary}

Given a dance sequence $\mathbf{P} \in \mathbb{R}^{T \times (J \times 3)}$, where $T$ is the time length and $J$ is joint amount. 
Using two pretrained Pose VQ-VAEs~\cite{siyao2022bailando} $\mathcal{F}^u_{VAE}$ and  $\mathcal{F}^l_{VAE}$ with codebook $\mathcal{Z}^u$ and $\mathcal{Z}^l$ for the upper and lower half bodies, it separately encodes upper and lower body movements into compositional code pairs $\mathbf{p}=[\mathbf{p}^u,\mathbf{p}^l]$:
\begin{equation}
    \mathbf{p}^u = \mathcal{F}^u_{VAE}(\mathbf{P}), \mathbf{p}^l = \mathcal{F}^l_{VAE}(\mathbf{P}),
\end{equation}
where $\mathbf{p}^u \in (\mathcal{Z}^u)^{T'}$, $\mathbf{p}^l \in (\mathcal{Z}^l)^{T'}$. Here $T^\prime=T/ \lambda$, $\lambda$ is the temporal down-sampling rate. The input music feature $\mathbf{m}\in \mathbb{R}^{T^\prime \times D_m}$, where $D_m$ is the channel dimension of the input music features. Then we embed music features $\mathbf{m}$, upper $p^u$ and lower $p^l$ pose codes to learnable features with three separate linear layers:
$\mathbf{X}_\mathrm{m}=\text{Linear}(\mathbf{m}), \mathbf{X}_\mathrm{u}=\text{Linear}(\mathbf{p}^u)$ and $\mathbf{X}_\mathrm{l}=\text{Linear}(\mathbf{p}^l),$
where $\mathbf{X}_\mathrm{m} \in \mathbb{R}^{T^\prime \times D}$, $\mathbf{X}_\mathrm{u} \in \mathbb{R}^{T^\prime \times D}$,  $\mathbf{X}_\mathrm{l} \in \mathbb{R}^{T^\prime \times D}$ and D is the channel dimension of the features.

The dance generation task is reframed as selecting the most probable future pose code from the codebook $\mathcal{Z}$, conditioned on the music and prior movements. Since the upper and lower bodies are modeled separately, maintaining coherence and avoiding asynchrony (e.g., opposing directions) requires cross-conditioned prediction, leveraging mutual information between existing movements:
\begin{align}
    \hat{p}_{t}^{u}=\arg \max _{k} \mathbb{P}\left(\mathbf{z}_{k}^{u} \mid \mathbf{m}_{1 \ldots t}, p_{0 \ldots t-1}^{u}, p_{0 \ldots t-1}^{l}\right), \\
    \hat{p}_{t}^{l}=\arg \max _{k} \mathbb{P}\left(\mathbf{z}_{k}^{l} \mid \mathbf{m}_{1 \ldots t}, p_{0 \ldots t-1}^{u}, p_{0 \ldots t-1}^{l}\right).
\label{eq:predict}
\end{align}
At each time step $t$, MotionGPT~\cite{siyao2022bailando} estimates the probabilities of pose codes $\mathrm{z}_i \in \mathbb{Z}$ and selects the one with the most probable pose codes as the predicted upper pose $\hat{p}_t^{u}$ and predicted lower pose $\hat{p}_t^{l}$.

Specifically, the input feature $\mathbf{X}_i \in \mathbb{R}^{(3 \times T ^ \prime) \times D}$ are concatenated by upper body motion feature, lower body motion feature and music feature as follows:
$\mathbf{X}_i = \operatorname{Concat}(\mathbf{X}_\mathrm{m}, \mathbf{X}_\mathrm{u}, \mathbf{X}_\mathrm{l}).$
Subsequently, $\mathbf{X}_i$ is passed through successive Autoregressive Transformer layers $\mathcal{F}_{AR}$,  followed by a linear transformation and a softmax layer, generates the probability distributions for predicting body pose collectively:
\begin{equation}
\mathbf{a}^{h}=\text{Softmax}(\text{Linear}(\mathcal{F}_{AR}(\mathbf{X}_i)),
\end{equation}
where $\mathbf{a}^{h}= [\mathbf{a}^u, \mathbf{a}^l]$, $\mathbf{a}^l$ denotes  that of upper body and $\mathbf{a}^u$ denotes that of lower body.

Finally, the model is optimized via supervised training with the cross-entropy loss on action probability $\mathrm{a}_{t}^{h}$ at each time step $t$ and for both upper body and lower body:
\begin{equation}
\mathcal{L}_{CE}=\frac{1}{T^{\prime}} \sum_{t=0}^{T^{\prime}-1} \sum_{h = u, l} \text { CrossEntropy }\left(\mathrm{a}_{t}^{h}, p_{t+1}^{h}\right) .
\end{equation}
In inference, the MotionGPT predicts pose codes according to the initial pose code and the entire music, followed by Pose VQ-VAE decoders generating a new dance.

\subsection{Phase-Based Rhythm Feature Extraction}
\label{sec:Rhythm Feature Extraction}

\begin{figure}[h!]
    \centering
    \includegraphics[width=1\linewidth]{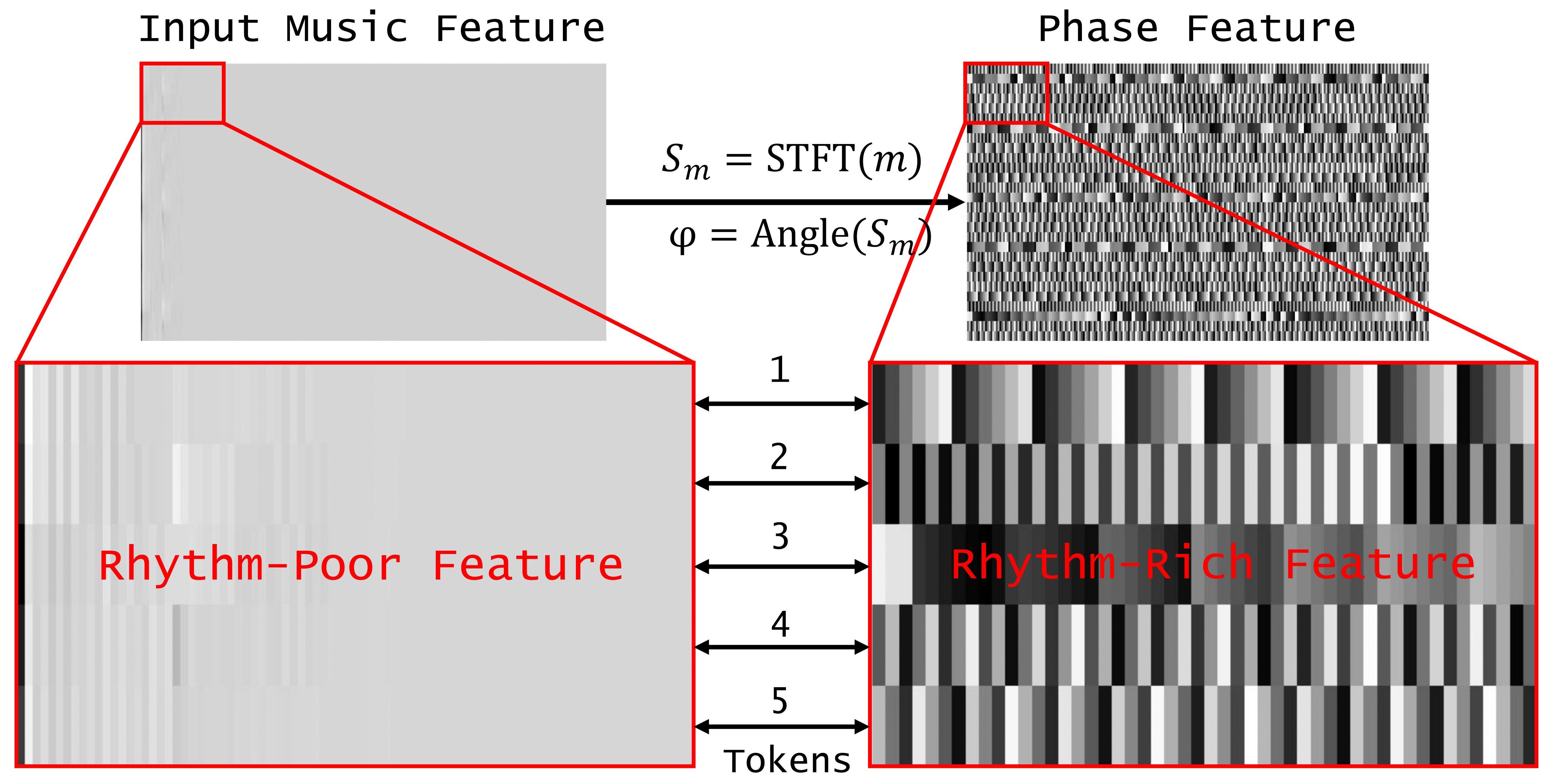}
    \caption{Visualizing the input music features and phase features. We notice that only a small portion of the input music features contains meaningful musical information, i.e., rhythm-poor features. And after a short-time Fourier transform (STFT), each corresponding token has a rich periodic Phase Feature, i.e., rhythm-rich features.}
    \vspace{-12pt}
    \label{fig:STFT}
\end{figure}

To enhance rhythmic feature extraction in music-driven dance generation, we propose Phase-Based Rhythm Extraction (PRE), which decouples rhythm from musical semantics by leveraging the inherent periodicity and temporal positioning of phase representations. The architecture of PRE is shown in the green part of Figure~\ref{fig:pipeline}. Unlike existing methods that entangle rhythm with melody and harmony, PRE enables independent and precise rhythm modeling as shown in Figure~\ref{fig:STFT}, forming a solid foundation for gating-enhanced rhythm-aware mechanism.

Specifically, we adopt the Short-Time Fourier Transform (STFT) instead of the conventional Fourier Transform (FT), since the non-stationary nature of music and the sliding-window approach of STFT effectively captures sudden tempo shifts and complex rhythmic patterns~\cite{{welker2022speech}}.
Specifically, we first perform STFT on the input music feature to extract phase angles:
\begin{equation}
\begin{aligned}
\mathbf{S}_{m} & =\operatorname{STFT}\left(\mathrm{\mathbf{m}}\right), \\
\boldsymbol{\varphi} & =\operatorname{Angle}\left(\mathbf{S}_{m}\right),
\end{aligned}
\end{equation}
where $\mathbf{S}_m$ represents the complex spectrogram obtained via $\operatorname{STFT}
$, $\operatorname{Angle}(\cdot)$ computes the phase angle of a complex-valued input and $\boldsymbol{\varphi}'$ denotes the phase angles extracted from $\mathbf{S}_m$.
By taking the angle of the STFT output, we obtain the music phase data $\boldsymbol{\varphi}' \in \mathbb{C}^{T_\varphi\times D_{\varphi}}$, where $T_\varphi$ represents the number of time frames obtained after the STFT transformation, and $D_{\varphi}$ is the channel dimension of the $\boldsymbol{\varphi}'$.

The extracted phase angles $\boldsymbol{\varphi}'$ inherently reflect temporal shifts and periodic structures of musical rhythms: low-frequency phase variations correspond to broader rhythmic patterns, while high-frequency variations capture finer rhythmic details.
We also apply a center-cropping strategy, selectively focusing on the most informative phase regions:
\begin{equation}
    \boldsymbol{\varphi} = \text{CenterCrop}(\boldsymbol{\varphi}'),
\end{equation} 
where $\boldsymbol{\varphi} \in \mathbb{C}^{T'\times D_{\varphi}}$ has the same temporal length $T'$, addressing the potential temporal misalignment  and preserving critical rhythmic cues.

Subsequently, the phase features through a linear transformation, enhancing phase features representing musical rhythms. The process is shown in the following equation:
\begin{equation}
    \mathbf{X}_{\varphi} = \text{ReLU} \left( \text{BN} \left( \text{Linear} (\varphi) \right) \right),
\end{equation} 
where the embedded phase features $\mathbf{X}_{\varphi} \in \mathbb{C}^{T^\prime \times D}$ encode rich rhythmic information, providing a clearer and more structured representation of rhythmic and musical elements.
Finally, we fuse the input feature $\mathbf{X}_i$ with the rhythmic features $\mathbf{X}_\varphi$ by concatenating them:
$\mathbf{X}_\gamma = \operatorname{Concat}(\mathbf{X}_{\varphi}, \mathbf{X}_{\varphi}, \mathbf{X}_{\varphi}).$
Then, we sum the rhythm-enhanced features with the original music, upper body, and lower body motion features: $
    \mathbf{X} = \mathbf{X}_f + \mathbf{X}_\gamma,$ 
where $\mathbf{X}$ represents the rhythmically enhanced features.

\subsection{Global Beat Attention via Temporal Gating}
\label{sec:TGCA}

\begin{figure}[h]
    \centering
    \includegraphics[width=1\linewidth]{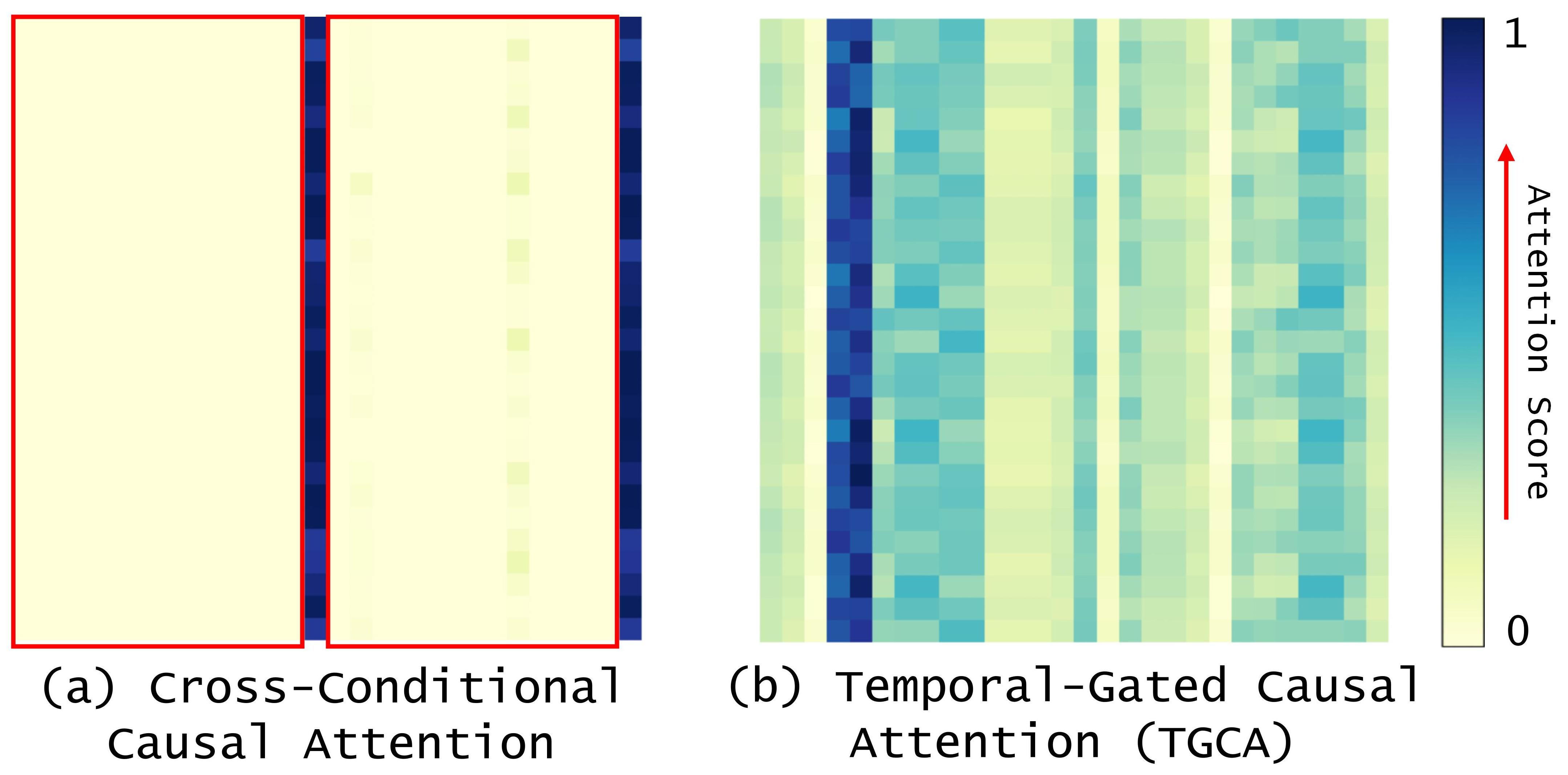}
    \caption{Heat map of attention. The two displayed images show the results learned by Cross-Conditional Causal Attention (a) and Temporal-Gated Causal Attention (b) under the same piece of music. The brighter the color, the higher the attention; the darker the color, the lower the attention. TGCA the next token has a clear control signal, preventing overly random dance movements and achieving enhanced rhythm-aware feature representation.}
    \label{fig:GCCA}
\end{figure}

Leveraging these rhythm-rich features, we next investigate how to integrate them into the alignment process between music and dance.  However, prior methods~\cite{siyao2022bailando, zhuang2023gtn, siyao2023bailando++, huang2024enhancing, wang2024explore}  relying on cross-conditional causal attention mechanism~(denoted as $\operatorname{C^3Attention}$) exhibit a critical misalignment issue.
As shown in Figure~\ref{fig:GCCA} (a), we visualize the cross-conditional causal attention mechanism and observe that the next predicted token lacks direct control signals from historical tokens, failing to establish a stable global beat attention. This misalignment leads to error accumulation, which we identify as a key factor behind the unnatural dance generation in prior work.

Building on this, we propose Temporal Gated Causal Attention (TGCA) to enhance global rhythm feature attention, improving music-dance alignment and ensuring the next token is guided by explicit historical information, preventing random action generation. As illustrated in the blue section of Figure~\ref{fig:pipeline}, the rhythmically enhanced features $\mathbf{X}$ are passed through TGCA module, which can be expressed as the following:
\begin{equation}
    \text{Gating}(\mathbf{X})=\text{SiLU}(\text{Linear}(\mathbf{X})).
\end{equation}
The final output is obtained by the element-wise multiplication of cross-conditional causal attention $\operatorname{C^3Attention}$ and $\operatorname{Gating}$, as shown in the following equation:
\begin{equation}
\begin{aligned}
&\mathbf{X}_{attn}=\text{TGCA}(\mathbf{X})\\
=\text{C}^3&\text{Attention}(\mathbf{X}) \odot \text{Gating}(\mathbf{X}).
\end{aligned}
\end{equation}
As shown in Figure~\ref{fig:GCCA} (b), our proposed gated causal attention mechanism allows historical tokens to better predict the next token, implying an enhanced influence of global rhythmic patterns on dance generation.

\subsection{Mamba-Based Parallel Motion Modeling}
\label{sec:Mamba-Based Motion Modeling}

After TGCA enhances dance movements in $\mathbf{X}_{attn}$ with global rhythmic perception, we looking into dance modeling for generation.  Dance generation faces two key challenges: (1) Dance is a complex spatial-temporal sequence encoding trajectory, posture, and velocity over time, requiring a model capable of capturing both local transitions and global rhythmic variations. (2) The upper and lower body exhibit distinct movement patterns and speeds, necessitating separate processing to maintain coherence and diversity. 
To address these issues, we propose a parallel Mamba motion modeling architecture that separately models music, upper-body, and lower-body movements while leveraging state space modeling to ensure fluid and diverse motion generation.

Specifically, to ensure each action is primarily influenced by relevant preceding movements, we introduce a GateMlp in the Mamba block, as shown in Figure~\ref{fig:pipeline}. This mechanism selectively retains essential motion features while filtering out less relevant information, thereby enhancing the coherence and expressiveness of the generated dance sequences. The process is formulated as follows:
\begin{equation}
    \mathbf{X}_{mb}' = \text{Mamba}(\text{RMSNorm}(\mathbf{X}_{attn})) + \mathbf{X}_{attn},
\end{equation}
\vspace{-25pt}
\begin{equation}
    \mathbf{X}_{mb} = \text{GateMlp}(\text{RMSNorm}(\mathbf{X}_{mb}')) + \mathbf{X}_{mb}'.
\end{equation}
The gating mechanism dynamically regulates the frequency, speed, and amplitude of movements in response to musical beats, ensuring that dance sequences remain synchronized with the music. Finally, three parallel Mamba architectures model upper- and lower-body dance sequences separately, therefore ensure the generated dance movements are expressive, diverse, and temporally coherent.

\begin{table*}[htbp]
\setlength{\tabcolsep}{4pt}
\renewcommand{\arraystretch}{0.95}
\caption{Comparison with state-of-the-art methods on the AIST++ dataset.
\textnormal{Underlining indicates the best performance among existing methods. \textcolor{cyan}{Blue} indicates results that surpass the best existing method. $\downarrow$ indicates that lower values are better, while $\uparrow$ indicates that higher values are better.
`$\ast$' represents the dances generated by ``Li \textit{et al.}” that exhibit significant jitteriness, resulting in extremely high velocity variation, as also noted in~\cite{li2021ai}. `$\dagger$' indicates the reproduced results obtained using Bailando++ official code and published checkpoints.
`$\ddagger$' denotes the results that are averaged over five independent training runs.}}
\label{tab:AIST++}
\vspace{-10pt}
\renewcommand{\arraystretch}{0.9}
\begin{center}
\begin{small}
\begin{sc}
\begin{tabular}{lccccccc}
\toprule
\multicolumn{1}{c}{\multirow{2}{*}{\textbf{Method}}} 
& \multicolumn{1}{c}{\multirow{2}{*}{\textbf{Venue}}} 
& \multicolumn{2}{c}{\textbf{Motion Quality}} 
& & \multicolumn{2}{c}{\textbf{Motion Diversity}} 
& \multirow{2}{*}{\textbf{Beat Align Score$\uparrow$}} \\ 
\cline{3-4} \cline{6-7}
& & $\text{FID}_k\downarrow$ & $\text{FID}_g\downarrow$ & & $\text{Div}_k\uparrow$ & $\text{Div}_g\uparrow$ \\
\midrule
Ground Truth        & -- & 17.10 & 10.60 & & 8.19 & 7.45 & 0.2374 \\ 
\midrule
Li \textit{et al.} \cite{li2020learning} & Arxiv 2020 & 86.43 & 43.46 & & $^{\ast} \text{6.85}$ & 3.32 & 0.1607 \\
DanceRevolution \cite{huang2021}   & ICLR 2021 & 73.42 & 25.92 & & 3.52 & 4.87 & 0.1950 \\ 
DanceNet \cite{10.1145/3485664}          & TOMM 2022 & 69.18 & 25.49 & & 2.86  & 2.85 & 0.1430 \\ 
FACT  \cite{li2021ai}             & ICCV 2021 & 35.35 & 22.11 & & 5.94 & 6.18 & 0.2209 \\ 
Bailando \cite{siyao2022bailando}      & CVPR 2022 & 28.16 & 9.62  & & 7.83 & 6.34 & 0.2332 \\
EDGE \cite{tseng2023edge}              & CVPR 2023 & 42.16 & 22.12 & & 3.96 & 4.61 & 0.2334 \\
DiffDance \cite{10.1145/3581783.3612307}         & ACMMM 2023 & 24.09 & 20.68  & & 6.02 & 2.89 & 0.2418 \\
$^{\dagger} \text{Bailando++}$ \cite{siyao2023bailando++}        &  TPAMI 2023 & \underline{22.74} & 11.58  & & 7.94 & 6.34 & 0.2263 \\
E3D2~\cite{wang2024explore} & AAAI 2024  & 26.25 & \underline{8.94}  & & \underline{7.96} & \underline{6.49} & 0.2232 \\
Lodge \cite{li2024lodge}             & CVPR 2024 & 37.09 & 18.79 & & 5.58 & 4.85 & \underline{0.2423} \\ 
\midrule
\rowcolor{gray!20}Ours (Best)        & & \textbf{11.67}\textbf{(\textcolor{cyan}{-11.07})} & 11.90 & & \textbf{8.52}\textbf{(\textcolor{cyan}{+0.58})}  & \textbf{7.55}\textbf{(\textcolor{cyan}{+1.21})} &\textbf{0.2714}\textbf{(\textcolor{cyan}{+2.91\%})} \\
\rowcolor{gray!20}$^{\ddagger} \text{Ours (Average)}$    & & \textbf{15.48}\textbf{(\textcolor{cyan}{-7.26})} & 13.57 & & 7.85  & \textbf{7.49}\textbf{(\textcolor{cyan}{+1.15})} &\textbf{0.2779}\textbf{(\textcolor{cyan}{+3.56\%})} \\
\bottomrule
\end{tabular}
\vspace{-10pt}
\end{sc}
\end{small}
\end{center}
\end{table*}

\begin{figure*}[htbp]
    \centering
    \includegraphics[width=0.9\linewidth]{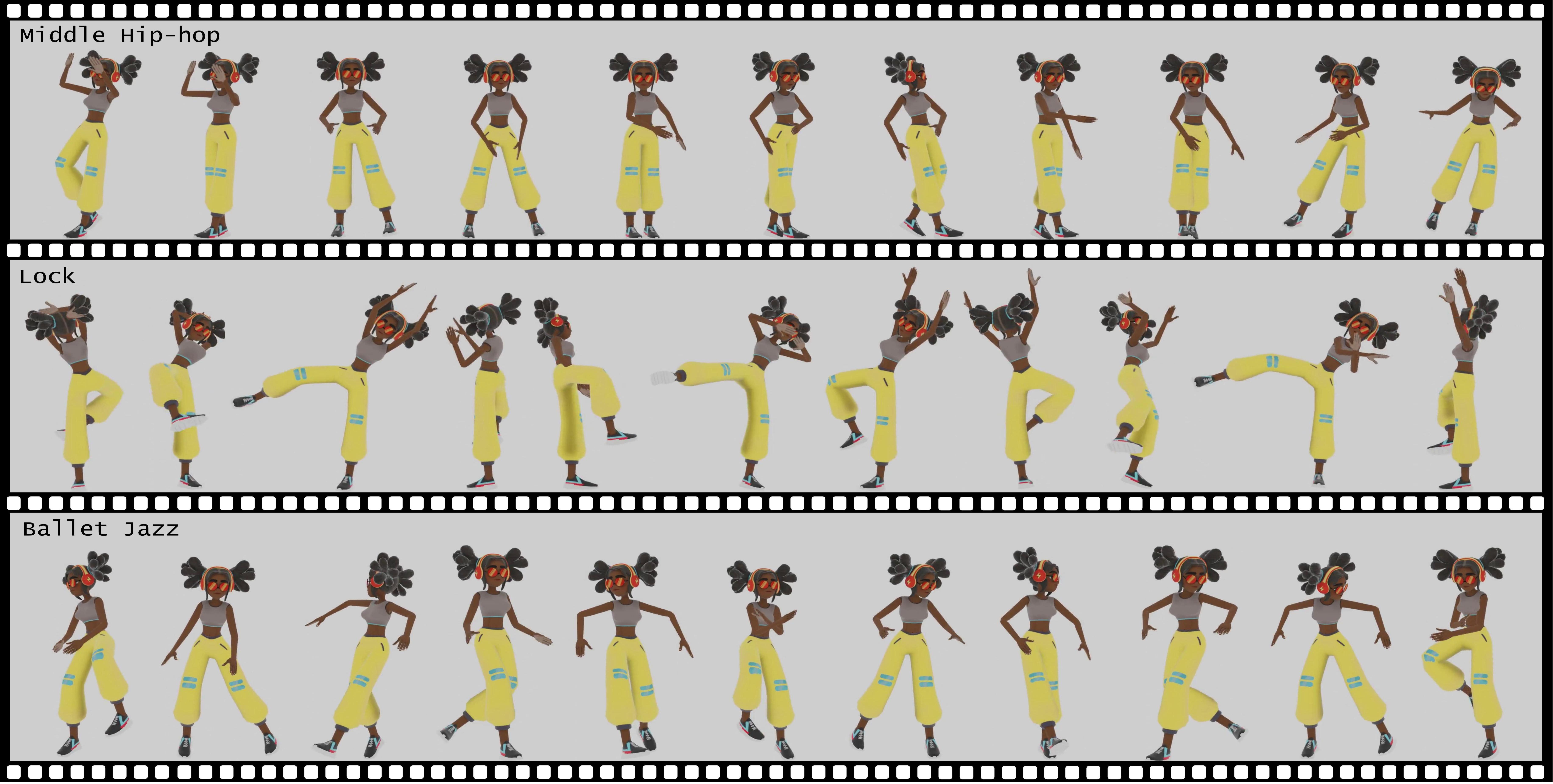}
    \caption{Dance examples generated by our proposed method on various types of music. The demo videos can be found in the supplementary material. The character is sourced from Mixamo~\cite{Mixamo}.} 
    \vspace{-8pt}
    \label{fig:4}
\end{figure*}

\section{Experiments}
\label{sec:Experiments}
\subsection{Experimental Setup}
\noindent\textbf{Dataset:}
We follow prior works~\cite{siyao2022bailando, 10.1145/3581783.3612307, wang2024explore} and perform all experiments on the AIST++ dataset~\cite{li2021ai}, the most widely used benchmark for music-driven dance generation that avoids observable biases (e.g., offensive postures). AIST++ consists of 991 high-quality 3D pose sequences recorded at 60 FPS in skinned multi-person linear (SMPL) format~\cite{10.1145/2816795.2818013}, with 951 sequences designated for training and 40 reserved for evaluation. Following~\cite{siyao2022bailando, 10.1145/3581783.3612307, wang2024explore}, we generate 40 pieces of dance sequences on the test set of AIST++ dataset, and sample the generated dance sequence with length of 20 seconds for performance evaluation.

\noindent\textbf{Evaluation Metrics:} 
We follow previous studies~\cite{li2021ai, siyao2022bailando} to quantitatively evaluate the generated samples in three key aspects: quality, diversity, and beat-dance alignment. Prior to evaluation, it is necessary to extract kinetic features~\cite{Onuma2008FMDistanceAF} and geometric features~\cite{10.1145/1186822.1073247} from both the ground truth and the generated samples via a specific toolbox~\cite{gopinath2020fairmotion}. For dance quality, we compute two types of the Fréchet Inception Distance ($\text{FID}$)~\cite{10.5555/3295222.3295408}, including $\text{FID}_{k}$ based on kinetic features and $\text{FID}_{g}$ based on geometric features.
Lower $\text{FID}_k$ indicates that the generated dance has more similar kinetic features to the original dance, meaning the generated dances are more natural and smooth.
Lower $\text{FID}_g$ indicates that the generated dance has more similar geometric features, reflecting the quality of choreography. For dance diversity, we employ metrics $\text{Div}_{k}$ and $\text{Div}_{g}$~\cite{li2021ai}, which denote the distance of kinetic and distance of geometric features, respectively. Higher $\text{Div}_k$ and $\text{Div}_g$ signify the greater amplitude and diversity of motion, respectively. Finally, to evaluate beat-dance alignment, we use the Beat Align Score ($\text{BAS}$)~\cite{li2021ai}, which quantifies the average distance between a music beat and its nearest dance beat.  Higher $\text{BAS}$ indicates greater degree of beat-dance alignment.

\subsection{Comparison with State-of-the-Art Methods }
To validate the superiority of our method, we conduct comparative  experiments with the state-of-the-art methods~\cite{li2020learning, huang2021, 10.1145/3485664, li2021ai, siyao2022bailando, tseng2023edge, 10.1145/3581783.3612307, siyao2023bailando++, wang2024explore, li2024lodge}, as well as the ground truth. Following~\cite{siyao2022bailando, 10.1145/3581783.3612307, wang2024explore}, we generate 40 pieces of dance sequences on the AIST++ test set, and sample the generated dance sequence with the length of 20 seconds to compute the metrics mentioned above.  Among the compared methods, ``Li \textit{et al.}"~\cite{li2020learning}, DanceNet~\cite{10.1145/3485664} and FACT~\cite{li2021ai} use the results from the AIST++ benchmark~\cite{li2021ai}, while the results of DanceRevolution~\cite{huang2021} are derived from its reproduction as presented in Bailando~\cite{siyao2022bailando}. EDGE~\cite{tseng2023edge} utilizes  the reproduced results from the Lodge~\cite{li2024lodge}. The Bailando, DiffDance~\cite{10.1145/3581783.3612307} and Lodge methods use the experimental results from their respective publications.

The comparative results are presented in Table~\ref{tab:AIST++}. Our method achieves state-of-the-art performance, significantly outperforming existing methods in terms of $\text{FID}_k$, $\text{Div}_g$, $\text{Div}_k$, and Beat Align Score ($\text{BAS}$), achieving the best results. Additionally, our method ranks third on the $\text{FID}_g$ metric. The generated dance examples of our proposed Danceba on various types of music are shown in Figure~\ref{fig:4}, and the demo videos are provided in the supplementary files.

\noindent\textbf{Motion Quality Metrics:} 
As illustrated in Table~\ref{tab:AIST++}, our method outperforms existing methods in overall quality performance.
To be specific, our method improves on $\text{FID}_k$ by 48.68\% (11.67 vs. 22.74), while exhibits competitive edge of $\text{FID}_g$.
Further analysis of the employed quality metrics reveals distinct roles for the two metrics types. The metric $\text{FID}_k$, defined based on velocity and energy, primarily captures the kinetic characteristics of the generated dance, while $\text{FID}_g$, derived from multiple manually designed motion templates, emphasizes the geometric structural properties of the dance. Thus, our method demonstrates superior performance in modeling the kinetic aspects of sequential motions, as evidenced by its exceptional $\text{FID}_k$ scores. Although it exhibits a modest gap in $\text{FID}_g$ (11.90 vs. 11.58) compared to the state-of-the-art Bailando++~\cite{siyao2023bailando++}, it achieves a 19.0\% higher $\text{Div}_g$ (7.55 vs. 6.34), indicating improved motion diversity with comparable generative quality. Despite this minor trade-off, our method outperforms existing approaches overall, achieving a robust balance of efficiency and quality across a broader range of metrics.

\begin{figure}[t]
    \centering
    \includegraphics[width=\linewidth]{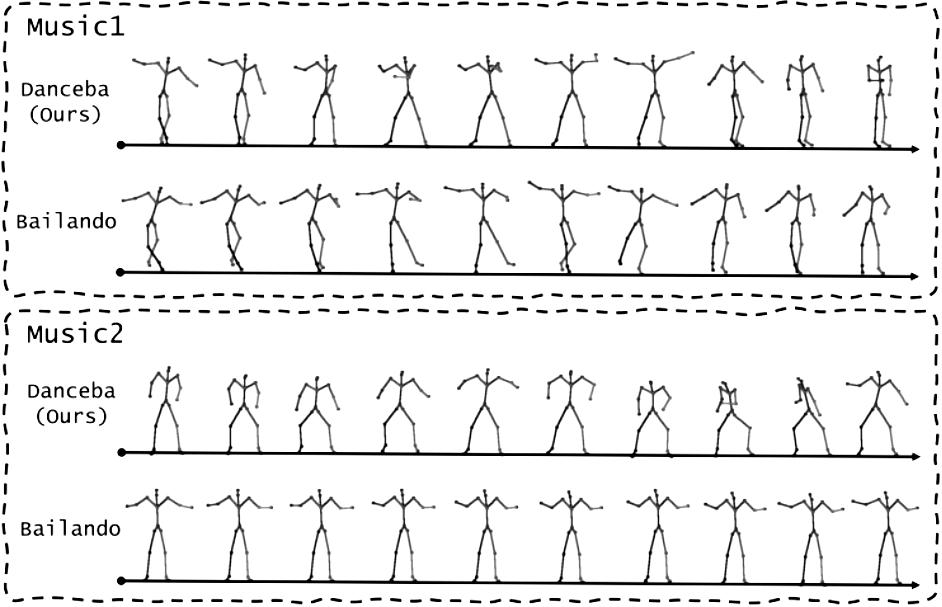}
    \caption{Comparison with the state-of-the-art method Bailando~\cite{siyao2022bailando}. Visual comparisons with Bailando can be found in the supplementary video.}
    \vspace{-14pt}
    \label{fig:compare_demo}
\end{figure}

\noindent\textbf{Motion Diversity Metrics:} 
Regarding diversity evaluation, Danceba achieves the best diversity performance, which improves $\text{Div}_k$ by 7.0\% (8.52 vs. 7.96) and even slightly outperforms the Ground Truth (8.52 vs. 8.19). Meanwhile, our method improves $\text{Div}_g$ by 16.3\% (7.55 vs. 6.49) and also slightly surpasses the Ground Truth (7.55 vs. 7.45), as illustrate in Table~\ref{tab:AIST++}.
For a visual comparison, we also present a comparison with the state-of-the-art method Bailando, as illustrated in Figure~\ref{fig:compare_demo}. It can be observed that Bailando exhibits relatively small movement amplitudes, with some dance movements missing and the actions appearing stiff. In contrast, our method shows larger movement amplitudes and movement diversity.

\begin{figure}[htbp]
    \centering
    \renewcommand{\arraystretch}{1}
    \includegraphics[width=0.98\linewidth]{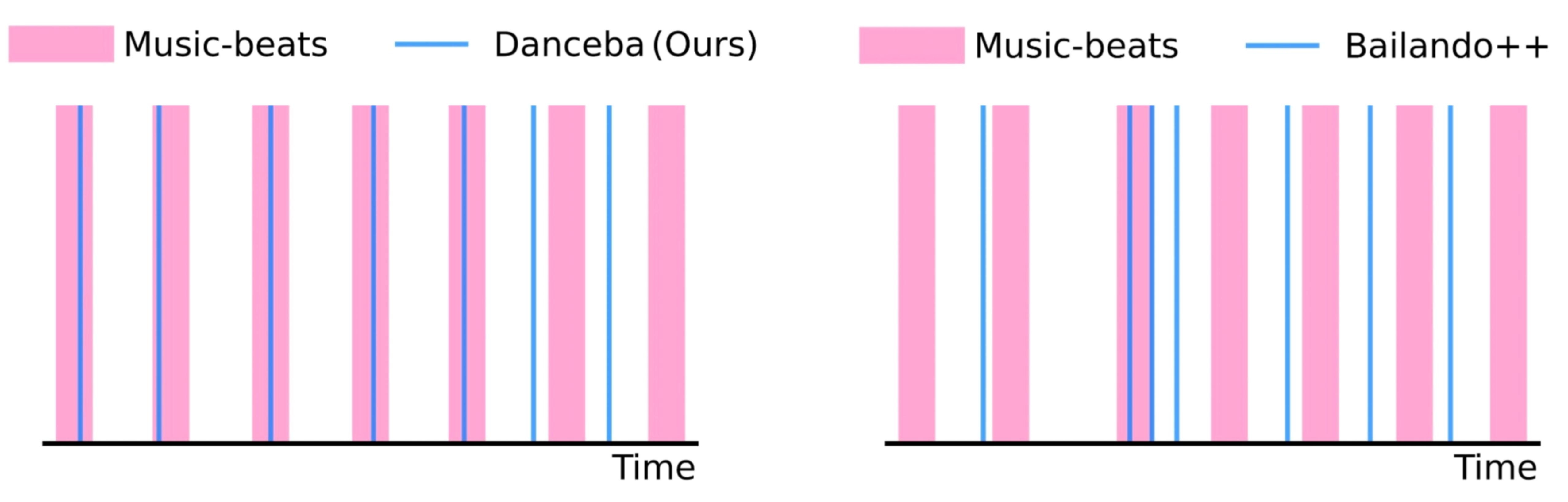}
    \caption{Beats alignment visualization, where the horizontal axis shows frame indices of beat events. Comparing Danceba with Bailando++~\cite{siyao2023bailando++}, we can find that the distance between motion beats and music beats generated by our method is smaller. This indicating that Danceba performs better in terms of rhythmic alignment.}
    \vspace{-12pt}
    \label{fig:BAS}
\end{figure}

\noindent\textbf{Beat-dance Alignment Metrics:}
Danceba achieves the best performance in beat-dance alignment, as quantified by the BAS metric. Specifically, it improves BAS by 12.0\% (0.2714 vs. 0.2423) compared to the previous best result and surpasses the ground truth by 14.3\% (0.2714 vs. 0.2374), as shown in Table~\ref{tab:AIST++}. Figure~\ref{fig:BAS} further visualizes the alignment between music beats and motion beats of the generated dance compared to Bailando++~\cite{siyao2023bailando++}. The results show that Danceba achieves a significantly smaller beat distance, demonstrating its ability to generate dance sequences that more accurately follow musical rhythms.

\subsection{Ablation Studies}
\label{sec:Ablation}

We conduct ablation studies to evaluate the contributions of the key modules (i.e., PRE, TGCA, and PMMM) in Danceba, where ``w/o" denotes the removal of a specific module. ``w/o PMMM” refers to the setting where PMMM is replaced with TGCA layers to ensure fairness. The results are presented in Table~\ref{tab:ablation1}. Additionally, we assess our method using a single Mamba structure instead of the parallel Mamba structure for motion modeling, referred to as Danceba-Single, with results shown in Table~\ref{tab:ablation_mamba}.

Note that, all ablation results are averaged over five independent training runs to ensure the credibility and robustness of our results.  The motion quality score ($\text{FID}$) is computed as the average of $\text{FID}_k$ and $\text{FID}_g$, while the motion diversity score ($\text{Div}$) is the average of $\text{Div}_k$ and $\text{Div}_g$.
\begin{table}[h]
\renewcommand{\arraystretch}{0.95}
\setlength{\tabcolsep}{8pt}
\caption{Ablation study of three key modules (i.e., PRE, TGCA, and PMMM), and ``w/o" indicates Danceba 
without specified module. Visual comparisons with Bailando can be found in the supplementary video.}
\label{tab:ablation1}
\vspace{-6pt}
\centering
\begin{small}
\begin{sc}
\begin{tabular}{l|ccc}
     \toprule
     Method & ${\text{FID}} \downarrow$ & ${\text{Div}} \uparrow$  & $\text{BAS} \uparrow$ \\
     \midrule
     Ground Truth & 13.85 & 7.82 & 0.2374 \\
     \midrule
     w/o TGCA & 33.22 & 6.74 & -- \\
     w/o PMMM & 25.24 & 5.78 & -- \\
     w/o PRE & 20.62 & 6.53 & 0.2474 \\
     \rowcolor{gray!20}Danceba & 14.53 & 7.67 & 0.2779 \\
     \bottomrule
\end{tabular}
\end{sc}
\end{small}
\end{table}

\noindent\textbf{Phase-Based Rhythm Extraction:} 
To validate the effectiveness of PRE, we replace the rhythm feature extraction module with a learnable linear layer as  used in previous works~\cite{siyao2022bailando, zhuang2023gtn, siyao2023bailando++, huang2024enhancing, wang2024explore}. Consequently, the motion quality (i.e., FID), motion diversity (i.e., Div), and beat alignment score (i.e., BAS) decrease by 6.09 (41.91\%), 1.14 (14.86\%), and 0.0305 (10.96\%), respectively, as illustrated in Table~\ref{tab:ablation1}. Specifically, this performance drop highlights  the contribution of PPE, which effectively  captures the critical rhythm features of music, a crucial aspect that previous methods have largely overlooked.

\noindent\textbf{Temporal-Gated Causal Attention:} 
As shown in Table~\ref{tab:ablation1}, we validate the effectiveness of the proposed TGCA by training a Danceba variant that replaces TGCA with conventional causal cross-conditional attention. The results demonstrate that the full Danceba model outperforms this variant in both FID and Div, highlighting TGCA’s ability to effectively leverage global rhythmic features extracted by PRE. This significantly enhances the contextual association between music and dance movements, resulting in dance sequences that not only accurately reflect musical rhythm but also exhibit greater coherence and expressive diversity.

\noindent\textbf{Parallel Mamba Motion Modeling:} 
To evaluate  the effectiveness of Parallel Mamba Motion Modeling, we train a model without  this module. As shown in Table \ref{tab:ablation1},  Danceba outperforms its variant without using PMMM in terms of both FID and Div, demonstrating the importance of parallel Mamba modeling. These results validate Mamba's superior sequential modeling capability, which is particularly well-suited for dance generation as it requires both fine-grained inter-modal interactions and long-sequence modeling.

To further assess the advantage of using a parallel Mamba architecture instead of a single Mamba architecture for motion modeling, we compare Danceba with Danceba-Single. Unlike Danceba-Single, which treats the body as a unified entity and disregards the distinction between upper and lower-body movements, PMMM explicitly models their distinct motion dynamics. Additionally, PMMM integrates rhythm-aware representation alongside these distinct pose features, enhancing beat-dance alignment with the musical rhythm. As shown in Table~\ref{tab:ablation_mamba}, Danceba significantly improves expressiveness and temporal coherence, with FID$_k$ and FID$_g$ scores improving by 67.02 and 60.40, respectively.

\begin{table}[t!]
\renewcommand{\arraystretch}{0.95}
\setlength{\tabcolsep}{4pt}
\caption{Ablation study on parallel Mamba motion modeling, where Danceba-Single denotes our method using single Mamba architecture.}
\label{tab:ablation_mamba}
\vspace{-6pt}
\centering
\begin{small}
\begin{sc}
\begin{tabular}{c|ccc}
\toprule
\textbf{Method} & $\text{FID}_k\downarrow$ & $\text{FID}_g\uparrow$  \\
\midrule
Ground Truth & 17.10 & 10.60 \\
\midrule
Danceba-Single & 82.50 & 83.97 \\
\rowcolor{gray!20}Danceba & 15.48\textbf{(\textcolor{cyan}{-67.02})} & 13.57\textbf{(\textcolor{cyan}{-60.40})} \\
\bottomrule
\end{tabular}
\vspace{-10pt}
\end{sc}
\end{small}
\end{table}

\section{Limitations}

Although Danceba achieves superior dance generation quality compared to state-of-the-art methods, it has two key limitations:
(1) Music feature encoding: Danceba uses a simple linear layer instead of a pre-trained audio model (e.g., Jukebox~\cite{dhariwal2020jukebox}, MERT~\cite{li2023mert}, or CLAP~\cite{elizalde2023clap}). While computationally efficient, this design may fail to capture nuanced musical structures, affecting rhythm-motion synchronization. Future work could explore pre-trained music encoders to enhance feature extraction. (2) 3D pose quantization: Danceba adopts the Pose VQ-VAE framework from Bailando~\cite{siyao2022bailando} for fair comparison with prior SOTA methods~\cite{siyao2022bailando, siyao2023bailando++, wang2024explore}. However, it has a limited encoding space, which may limit motion diversity. Such a trade-off between quantization efficiency and expressiveness could hinder  the preservation of fine-grained motion details. Exploring hierarchical quantization or adaptive codebook learning could further improve motion quality.

\section{Conclusion}

This paper has presented a novel framework Danceba for music-driven dance generation that enhances rhythm-aware feature representation and motion modeling. By introducing phase-based rhythm extraction and temporal-gated causal attention, our approach ensures precise beat-dance synchronization and effectively integrates rhythmic structures into dance movements. Additionally, parallel Mamba motion modeling separately captures the distinct motion dynamics of the upper and lower body while incorporating rhythm-aware representations, improving the naturalness and diversity of the generated dance. Extensive experiments demonstrate that Danceba significantly outperforms state-of-the-art methods, generating dance movements that are more rhythmic, natural, and diverse while maintaining strong alignment with musical beats. Future work could integrate pre-trained audio encoders for richer music representation, explore advanced quantization to enhance motion expressiveness and diversity, or extend to style-conditioned motion synthesis for broader applicability. 

\section*{Acknowledgments} 
This work was partly supported by the Fundamental Research Funds for the Central Universities (Grant No. 3072025YY0601), and the Beijing Nova Program (Grant No. 20230484261).

{
    \small
    \bibliographystyle{ieeenat_fullname}
    \bibliography{main}

\begin{thebibliography}{45}
\providecommand{\natexlab}[1]{#1}
\providecommand{\url}[1]{\texttt{#1}}
\expandafter\ifx\csname urlstyle\endcsname\relax
  \providecommand{\doi}[1]{doi: #1}\else
  \providecommand{\doi}{doi: \begingroup \urlstyle{rm}\Url}\fi

\bibitem[Alexanderson et~al.(2023)Alexanderson, Nagy, Beskow, and Henter]{alexanderson2023listen}
Simon Alexanderson, Rajmund Nagy, Jonas Beskow, and Gustav~Eje Henter.
\newblock Listen, denoise, action! audio-driven motion synthesis with diffusion models.
\newblock \emph{ACM Trans. Graph. (TOG)}, 2023.

\bibitem[Dao and Gu(2024)]{dao2024transformers}
Tri Dao and Albert Gu.
\newblock Transformers are {SSM}s: Generalized models and efficient algorithms through structured state space duality.
\newblock In \emph{Proc. Int. Conf. Mach. Learn. (ICML)}, 2024.

\bibitem[Dhariwal et~al.(2020)Dhariwal, Jun, Payne, Kim, Radford, and Sutskever]{dhariwal2020jukebox}
Prafulla Dhariwal, Heewoo Jun, Christine Payne, Jong~Wook Kim, Alec Radford, and Ilya Sutskever.
\newblock Jukebox: A generative model for music.
\newblock \emph{arXiv preprint arXiv:2005.00341}, 2020.

\bibitem[Dong et~al.(2024)Dong, Xue, Niu, Lan, Lu, Liu, and Qin]{MMD}
Kun Dong, Jian Xue, Zehai Niu, Xing Lan, Ke Lu, Qingyuan Liu, and Xiaoyu Qin.
\newblock Realistic full-body motion generation from sparse tracking with state space model.
\newblock In \emph{Proc. ACM Int. Conf. Multimedia (ACMMM)}, 2024.

\bibitem[Elizalde et~al.(2023)Elizalde, Deshmukh, Al~Ismail, and Wang]{elizalde2023clap}
Benjamin Elizalde, Soham Deshmukh, Mahmoud Al~Ismail, and Huaming Wang.
\newblock {CLAP} learning audio concepts from natural language supervision.
\newblock In \emph{Proc. IEEE Int. Conf. Acoust. Speech Signal Process. (ICASSP)}, 2023.

\bibitem[Gopinath and Won(2020)]{gopinath2020fairmotion}
Deepak Gopinath and Jungdam Won.
\newblock Fairmotion - tools to load, process and visualize motion capture data.
\newblock Github, 2020.

\bibitem[Gu and Dao(2023)]{gu2023mamba}
Albert Gu and Tri Dao.
\newblock Mamba: Linear-time sequence modeling with selective state spaces.
\newblock \emph{arXiv preprint arXiv:2312.00752}, 2023.

\bibitem[Heusel et~al.(2017)Heusel, Ramsauer, Unterthiner, Nessler, and Hochreiter]{10.5555/3295222.3295408}
Martin Heusel, Hubert Ramsauer, Thomas Unterthiner, Bernhard Nessler, and Sepp Hochreiter.
\newblock {GANs} trained by a two time-scale update rule converge to a local nash equilibrium.
\newblock In \emph{Proc. Adv. Neural Inf. Process. Syst. (NeurIPS)}, 2017.

\bibitem[Ho et~al.(2020)Ho, Jain, and Abbeel]{ho2020denoising}
Jonathan Ho, Ajay Jain, and Pieter Abbeel.
\newblock Denoising diffusion probabilistic models.
\newblock In \emph{Proc. Adv. Neural Inf. Process. Syst. (NeurIPS)}, 2020.

\bibitem[Holden et~al.(2016)Holden, Saito, and Komura]{holden2016deep}
Daniel Holden, Jun Saito, and Taku Komura.
\newblock A deep learning framework for character motion synthesis and editing.
\newblock \emph{ACM Trans. Graph. (TOG)}, 2016.

\bibitem[Huang et~al.(2024)Huang, He, Tang, Zhuang, Chen, Gao, Wu, Huang, and Meng]{huang2024enhancing}
Qiaochu Huang, Xu He, Boshi Tang, Haolin Zhuang, Liyang Chen, Shuochen Gao, Zhiyong Wu, Haozhi Huang, and Helen Meng.
\newblock Enhancing expressiveness in dance generation via integrating frequency and music style information.
\newblock In \emph{Proc. IEEE Int. Conf. Acoust. Speech Signal Process. (ICASSP)}, 2024.

\bibitem[Huang et~al.(2021)Huang, Hu, Wu, Sawada, Zhang, and Jiang]{huang2021}
Ruozi Huang, Huang Hu, Wei Wu, Kei Sawada, Mi Zhang, and Daxin Jiang.
\newblock Dance revolution: Long-term dance generation with music via curriculum learning.
\newblock In \emph{Proc. Int. Conf. Learn. Represent. (ICLR)}, 2021.

\bibitem[Kim et~al.(2022)Kim, Oh, Kim, Tong, and Lee]{kim2022brand}
Jinwoo Kim, Heeseok Oh, Seongjean Kim, Hoseok Tong, and Sanghoon Lee.
\newblock A brand new dance partner: Music-conditioned pluralistic dancing controlled by multiple dance genres.
\newblock In \emph{Proc. IEEE/CVF Conf. Comput. Vis. Pattern Recognit. (CVPR)}, 2022.

\bibitem[Kyan et~al.(2015)Kyan, Sun, Li, Zhong, Muneesawang, Dong, Elder, and Guan]{10.1145/2735951}
Matthew Kyan, Guoyu Sun, Haiyan Li, Ling Zhong, Paisarn Muneesawang, Nan Dong, Bruce Elder, and Ling Guan.
\newblock An approach to ballet dance training through {MS Kinect} and visualization in a {CAVE} virtual reality environment.
\newblock \emph{ACM Trans. Intell. Syst. Technol. (TIST)}, 2015.

\bibitem[Lee et~al.(2019)Lee, Yang, Liu, Wang, Lu, Yang, and Kautz]{lee2019dancing}
Hsin-Ying Lee, Xiaodong Yang, Ming-Yu Liu, Ting-Chun Wang, Yu-Ding Lu, Ming-Hsuan Yang, and Jan Kautz.
\newblock Dancing to music.
\newblock In \emph{Proc. Adv. Neural Inf. Process. Syst. (NeurIPS)}, 2019.

\bibitem[Li et~al.(2024{\natexlab{a}})Li, Shu, Cui, Yao, and Tang]{FTMoMamba}
Chengjian Li, Xiangbo Shu, Qiongjie Cui, Yazhou Yao, and Jinhui Tang.
\newblock {FTMoMamba}: {Motion} generation with frequency and text state space models.
\newblock \emph{arXiv preprint arXiv:2411.17532}, 2024{\natexlab{a}}.

\bibitem[Li et~al.(2020)Li, Yin, Chu, Zhou, Wang, Fidler, and Li]{li2020learning}
Jiaman Li, Yihang Yin, Hang Chu, Yi Zhou, Tingwu Wang, Sanja Fidler, and Hao Li.
\newblock Learning to generate diverse dance motions with {Transformer}.
\newblock \emph{arXiv preprint arXiv:2008.08171}, 2020.

\bibitem[Li et~al.(2021)Li, Yang, Ross, and Kanazawa]{li2021ai}
Ruilong Li, Shan Yang, David~A Ross, and Angjoo Kanazawa.
\newblock Ai choreographer: Music conditioned {3D} dance generation with {AIST}++.
\newblock In \emph{Proc. IEEE/CVF Int. Conf. Comput. Vis. (ICCV)}, 2021.

\bibitem[Li et~al.(2023)Li, Zhao, Zhang, Su, Ren, Zhang, Tang, and Li]{li2023finedance}
Ronghui Li, Junfan Zhao, Yachao Zhang, Mingyang Su, Zeping Ren, Han Zhang, Yansong Tang, and Xiu Li.
\newblock Finedance: A fine-grained choreography dataset for {3D} full body dance generation.
\newblock In \emph{Proc. IEEE/CVF Int. Conf. Comput. Vis. (ICCV)}, 2023.

\bibitem[Li et~al.(2024{\natexlab{b}})Li, Zhang, Zhang, Zhang, Guo, Zhang, Liu, and Li]{li2024lodge}
Ronghui Li, YuXiang Zhang, Yachao Zhang, Hongwen Zhang, Jie Guo, Yan Zhang, Yebin Liu, and Xiu Li.
\newblock Lodge: A coarse to fine diffusion network for long dance generation guided by the characteristic dance primitives.
\newblock In \emph{Proc. IEEE/CVF Conf. Comput. Vis. Pattern Recognit. (CVPR)}, 2024{\natexlab{b}}.

\bibitem[Li et~al.(2024{\natexlab{c}})Li, Yuan, Zhang, Ma, Chen, Yin, Xiao, Lin, Ragni, Benetos, et~al.]{li2023mert}
Yizhi Li, Ruibin Yuan, Ge Zhang, Yinghao Ma, Xingran Chen, Hanzhi Yin, Chenghao Xiao, Chenghua Lin, Anton Ragni, Emmanouil Benetos, et~al.
\newblock {MERT}: {A}coustic music understanding model with large-scale self-supervised training.
\newblock In \emph{Proc. Int. Conf. Learn. Represent. (ICLR)}, 2024{\natexlab{c}}.

\bibitem[Loper et~al.(2015)Loper, Mahmood, Romero, Pons-Moll, and Black]{10.1145/2816795.2818013}
Matthew Loper, Naureen Mahmood, Javier Romero, Gerard Pons-Moll, and Michael~J. Black.
\newblock Smpl: a skinned multi-person linear model.
\newblock \emph{ACM Trans. Graph. (TOG)}, 2015.

\bibitem[Mixamo()]{Mixamo}
Mixamo.
\newblock \url{https://www.mixamo.com/}.

\bibitem[M\"{u}ller et~al.(2005)M\"{u}ller, R\"{o}der, and Clausen]{10.1145/1186822.1073247}
Meinard M\"{u}ller, Tido R\"{o}der, and Michael Clausen.
\newblock Efficient content-based retrieval of motion capture data.
\newblock In \emph{Proc. ACM Conf. SIGGRAPH}, 2005.

\bibitem[Nichol et~al.(2021)Nichol, Dhariwal, Ramesh, Shyam, Mishkin, McGrew, Sutskever, and Chen]{nichol2021glide}
Alex Nichol, Prafulla Dhariwal, Aditya Ramesh, Pranav Shyam, Pamela Mishkin, Bob McGrew, Ilya Sutskever, and Mark Chen.
\newblock {GLIDE}: {T}owards photorealistic image generation and editing with text-guided diffusion models.
\newblock \emph{arXiv preprint arXiv:2112.10741}, 2021.

\bibitem[Onuma et~al.(2008)Onuma, Faloutsos, and Hodgins]{Onuma2008FMDistanceAF}
Kensuke Onuma, Christos Faloutsos, and Jessica~K. Hodgins.
\newblock Fmdistance: A fast and effective distance function for motion capture data.
\newblock In \emph{Eurographics}, 2008.

\bibitem[Qi et~al.(2023)Qi, Zhuo, Zhang, Liao, Fang, Liu, and Yan]{10.1145/3581783.3612307}
Qiaosong Qi, Le Zhuo, Aixi Zhang, Yue Liao, Fei Fang, Si Liu, and Shuicheng Yan.
\newblock {DiffDance}: Cascaded human motion diffusion model for dance generation.
\newblock In \emph{Proc. ACM Int. Conf. Multimedia (ACMMM)}, 2023.

\bibitem[Qian et~al.(2024)Qian, Xiao, Wu, Yang, Li, Wang, Wang, Kou, and Zhang]{qian2024smcd}
Ziyun Qian, Zeyu Xiao, Zhenyi Wu, Dingkang Yang, Mingcheng Li, Shunli Wang, Shuaibing Wang, Dongliang Kou, and Lihua Zhang.
\newblock {SMCD}: High realism motion style transfer via {Mamba}-based diffusion.
\newblock \emph{arXiv preprint arXiv:2405.02844}, 2024.

\bibitem[Radford et~al.(2019)Radford, Wu, Child, Luan, Amodei, and Sutskever]{radford2019language}
Alec Radford, Jeff Wu, Rewon Child, David Luan, Dario Amodei, and Ilya Sutskever.
\newblock Language models are unsupervised multitask learners.
\newblock \emph{OpenAI blog}, 2019.

\bibitem[Siyao et~al.(2022)Siyao, Yu, Gu, Lin, Wang, Qian, Loy, and Liu]{siyao2022bailando}
Li Siyao, Weijiang Yu, Tianpei Gu, Chunze Lin, Quan Wang, Chen Qian, Chen~Change Loy, and Ziwei Liu.
\newblock Bailando: {3D} dance generation by actor-critic {GPT} with choreographic memory.
\newblock In \emph{Proc. IEEE/CVF Conf. Comput. Vis. Pattern Recognit. (CVPR)}, 2022.

\bibitem[Siyao et~al.(2023)Siyao, Yu, Gu, Lin, Wang, Qian, Loy, and Liu]{siyao2023bailando++}
Li Siyao, Weijiang Yu, Tianpei Gu, Chunze Lin, Quan Wang, Chen Qian, Chen~Change Loy, and Ziwei Liu.
\newblock Bailando++: {3D} dance {GPT} with choreographic memory.
\newblock \emph{IEEE Trans. Pattern Anal. Mach. Intell. (TPAMI)}, 2023.

\bibitem[Sun et~al.(2020)Sun, Wong, Cheng, Kankanhalli, Geng, and Li]{sun2020deepdance}
Guofei Sun, Yongkang Wong, Zhiyong Cheng, Mohan~S Kankanhalli, Weidong Geng, and Xiangdong Li.
\newblock Deepdance: music-to-dance motion choreography with adversarial learning.
\newblock \emph{IEEE Trans. Multimedia (TMM)}, 2020.

\bibitem[Tseng et~al.(2023)Tseng, Castellon, and Liu]{tseng2023edge}
Jonathan Tseng, Rodrigo Castellon, and Karen Liu.
\newblock Edge: Editable dance generation from music.
\newblock In \emph{Proc. IEEE/CVF Int. Conf. Comput. Vis. (ICCV)}, 2023.

\bibitem[Tsuchida(2024)]{tsuchida2024dance}
Shuhei Tsuchida.
\newblock Dance information processing: {C}omputational approaches for assisting dance composition.
\newblock \emph{New Gener. Comput.}, 2024.

\bibitem[Vaswani et~al.(2017)Vaswani, Shazeer, Parmar, Uszkoreit, Jones, Gomez, Kaiser, and Polosukhin]{vaswani2017attention}
Ashish Vaswani, Noam Shazeer, Niki Parmar, Jakob Uszkoreit, Llion Jones, Aidan~N Gomez, {\L}ukasz Kaiser, and Illia Polosukhin.
\newblock Attention is all you need.
\newblock \emph{Proc. Adv. Neural Inf. Process. Syst. (NeurIPS)}, 2017.

\bibitem[Wang et~al.(2024)Wang, Zhuang, Li, Zhang, Zhong, Chen, Yang, Tang, and Wu]{wang2024explore}
Zilin Wang, Haolin Zhuang, Lu Li, Yinmin Zhang, Junjie Zhong, Jun Chen, Yu Yang, Boshi Tang, and Zhiyong Wu.
\newblock Explore {3D} dance generation via reward model from automatically-ranked demonstrations.
\newblock In \emph{Proc. AAAI Conf. Artif. Intell. (AAAI)}, 2024.

\bibitem[Welker et~al.(2022)Welker, Richter, and Gerkmann]{welker2022speech}
Simon Welker, Julius Richter, and Timo Gerkmann.
\newblock Speech enhancement with score-based generative models in the complex {STFT} domain.
\newblock \emph{arXiv preprint arXiv:2203.17004}, 2022.

\bibitem[Yan et~al.(2015)Yan, Ding, Guan, Sun, Li, and Zhang]{10.1145/2702613.2732759}
Shuo Yan, Gangyi Ding, Zheng Guan, Ningxiao Sun, Hongsong Li, and Longfei Zhang.
\newblock {OutsideMe}: {A}ugmenting dancer's external self-image by using a mixed reality system.
\newblock In \emph{Proc. Annu. ACM Conf. Extended Abstr. Hum. Factors Comput. Syst. (CHI)}, 2015.

\bibitem[Zeng et~al.(2024)Zeng, Huang, Wu, and Zheng]{zeng2024light}
Ling-An Zeng, Guohong Huang, Gaojie Wu, and Wei-Shi Zheng.
\newblock Light-t2m: A lightweight and fast model for text-to-motion generation.
\newblock \emph{arXiv preprint arXiv:2412.11193}, 2024.

\bibitem[Zhang et~al.(2024)Zhang, Liu, Chen, chen, Reid, Hartley, Zhuang, and Tang]{InfiniMotion}
Zeyu Zhang, Akide Liu, Qi Chen, Feng chen, Reid Reid, Richard Hartley, Bohan Zhuang, and Hao Tang.
\newblock {InfiniMotion}: {Mamba} boosts memory in {T}ransformer for arbitrary long motion generation.
\newblock \emph{arXiv preprint arXiv:2407.10061}, 2024.

\bibitem[Zhang et~al.(2025)Zhang, Liu, Reid, Hartley, Zhuang, and Tang]{zhang2025motion}
Zeyu Zhang, Akide Liu, Ian Reid, Richard Hartley, Bohan Zhuang, and Hao Tang.
\newblock Motion {Mamba}: {Efficient} and long sequence motion generation.
\newblock In \emph{Proc. Springer Eur. Conf. Comput. Vis. (ECCV)}, 2025.

\bibitem[Zheng et~al.(2024)Zheng, Qin, and He]{zheng2024beat}
Chenxi Zheng, Jing Qin, and Shengfeng He.
\newblock {Beat-It}: {B}eat-synchronized multi-condition {3D} dance generation.
\newblock In \emph{Proc. Springer Eur. Conf. Comput. Vis. (ECCV)}, 2024.

\bibitem[Zhu et~al.(2023)Zhu, Ma, Ro, Ci, Zhang, Shi, Gao, Tian, and Wang]{zhu2023human}
Wentao Zhu, Xiaoxuan Ma, Dongwoo Ro, Hai Ci, Jinlu Zhang, Jiaxin Shi, Feng Gao, Qi Tian, and Yizhou Wang.
\newblock Human motion generation: A survey.
\newblock \emph{IEEE Trans. Pattern Anal. Mach. Intell. (TPAMI)}, 2023.

\bibitem[Zhuang et~al.(2023)Zhuang, Lei, Xiao, Li, Chen, Yang, Wu, Kang, and Meng]{zhuang2023gtn}
Haolin Zhuang, Shun Lei, Long Xiao, Weiqin Li, Liyang Chen, Sicheng Yang, Zhiyong Wu, Shiyin Kang, and Helen Meng.
\newblock {GTN-Bailando}: {Genre} consistent long-term {3D} dance generation based on pre-trained genre token network.
\newblock In \emph{Proc. IEEE Int. Conf. Acoust. Speech Signal Process. (ICASSP)}, 2023.

\bibitem[Zhuang et~al.(2022)Zhuang, Wang, Chai, Wang, Shao, and Xia]{10.1145/3485664}
Wenlin Zhuang, Congyi Wang, Jinxiang Chai, Yangang Wang, Ming Shao, and Siyu Xia.
\newblock {Music2Dance}: {DanceNet} for music-driven dance generation.
\newblock \emph{ACM Trans. Multimedia Comput. Commun. Appl. (TOMM)}, 2022.

\end{thebibliography}
}

\end{document}